\documentclass{PoS}
\usepackage[percent]{overpic}
\pdfoutput=1
\title{Production of D$_{\rm_{S}}$ meson in Au+Au collision at $\sqrt{s_{\rm{NN}}}$ = 200 GeV}

\ShortTitle{D$_{\rm_{S}}$ meson in Au+Au collision}

\author{\speaker{Md Nasim (for the STAR Collaboration)}\\
        University of California, Los Angeles\\
        E-mail: \email{mnasim2008@gmail.com}}

\abstract{ We present the invariant yield and elliptic flow of D$_{\rm_{S}}$ as a function of
transverse momentum in Au+Au collisions at $\sqrt{s_{\rm{NN}}}$ = 200 GeV. The nuclear modification factors of  D$_{\rm_{S}}$ are found to be systematically higher than those of K$^{0}_{\rm_{S}}$ . The ratio between the yields of strange and non-strange open charm mesons is shown.  We find that such a ratio in  0-40$\%$  central Au+Au collisions is higher than the fragmentation baseline.  Our measurement indicates a substantial enhancement of D$_{\rm_{S}}$ production in Au+Au collisions with respect to p+p collisions as compared to non-strange D mesons. The elliptic flow of D$_{\rm_{S}}$ is also measured and compared to that of D$^{0}$ as well as model calculations. }

\FullConference{Critical Point and Onset of Deconfinement - CPOD2017\\
		7-11 August, 2017\\
		The Wang Center, Stony Brook University, Stony Brook, NY}

\begin{document}

\section{Introduction}
The primary purpose of relativistic heavy-ion collisions at the Relativistic Heavy
Ion Collider (RHIC) is to create the QCD matter under high temperature and high
density - Quark-Gluon Plasma (QGP), and study its properties. Due to their large
masses, heavy quarks are produced on a short time scale in hard partonic
scatterings during the early stages of the nucleus-nucleus collisions and the
probability of thermal production in the QGP phase is expected to be small.
Therefore, they are considered a good probe for studying the QGP. Among all the
open charm mesons, the charm-strange meson, D$_{\rm_{S}}$, is particularly sensitive to the
charm quark hadronization in the hot nuclear medium because of its unique
valence quark composition. Theoretical calculations predict that the production of D$_{\rm_{S}}$ can be influenced by the
charm-quark recombination with strange quarks whose production is enhanced in
the deconfined matter~\cite{rapp}.   Like multi-strange hadrons, D$_{\rm_{S}}$ mesons have smaller hadronic interaction cross-sections compared to the
non-strange D mesons and are expected to freeze out early. Therefore,
the elliptic flow ($v_{2}$) of D$_{\rm_{S}}$ is considered a better measure of the partonic contribution to the charm
hadron $v_{2}$ than that of D$^{0}$ or D$^{\pm}$.
The Heavy Flavor Tracker (HFT) at STAR provides an opportunity for D$_{\rm_{S}}$ measurements by reconstructing displaced decay vertices.

\section{Data set and analysis details}
\begin{figure}[!ht]
\begin{center}
\begin{overpic}[scale=0.45]{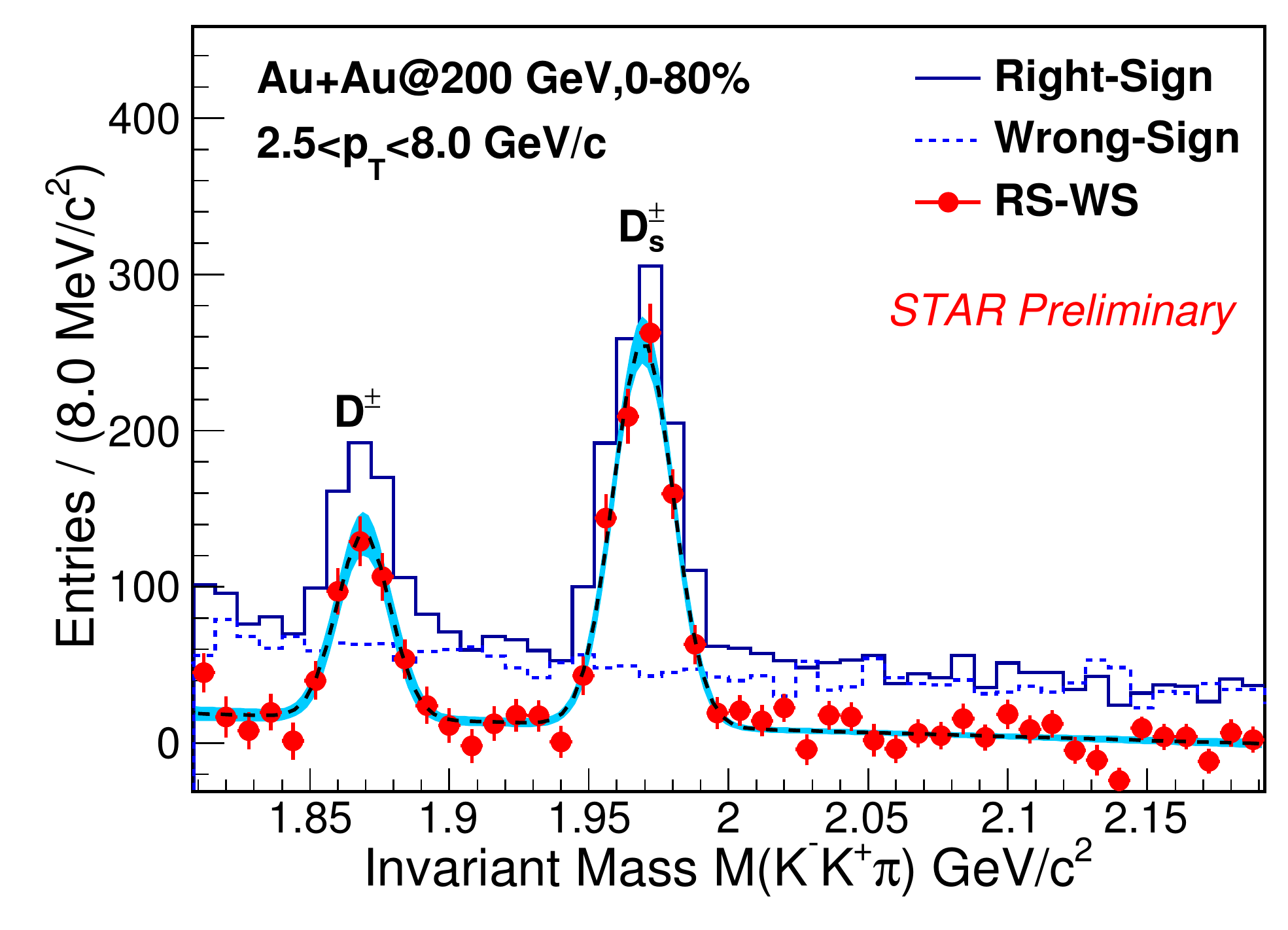}
\end{overpic}
\caption{(Color online)  The invariant mass distributions of $K^{+}K^{-}\pi$ triplets
  from Au+Au collisions at  \mbox{$\sqrt{s_{\rm{NN}}}$  = 200 GeV} for 0-80$\%$ centrality, 2.5 $<$ $p_{T}$ $<$
  8.0 GeV/$c$. }
\label{sig_08}
\end{center}
\end{figure}

The results presented here are based on an analysis of about 900 million minimum bias events taken
during the 2014 Au+Au run at $\sqrt{s_{\rm{NN}}}$ = 200 GeV. The Time Projection Chamber (TPC) and the Time-of-Flight
(TOF) detectors, both with full azimuthal coverage, are used for particle identification in the central rapidity (y) region ($|$y$|$ $<$ 1.0). The HFT detector is used to reconstruct the decay vertices. It is made of  three  layers, named as PiXeL detector (PXL),  Intermediate Silicon Tracker (IST) and Silicon Strip Detector (SSD).  A state-of-the-art thin 
Monolithic Active Pixel Sensors (MAPS) technology has been used in the PXL. There are  two layers of the MAPS  in the PXL,
which are placed at radii of 2.8 and 8 cm from the centre of the beam pipe, respectively. The track pointing resolution of the HFT detector is about 46 $\mu$m for 750 MeV/c kaons. We reconstruct D$_{\rm_{S}}$  through the decay channel, D$_{\rm_{S}}$$^{\pm}$  $\longrightarrow$ $\phi$ ($\phi$ $\longrightarrow$ $\it{K}^{+}$ + $\it{K}^{-}$) + $\pi^{\pm}$.
Topological and kinematic cuts are applied to reduce  the combinatorial background. The wrong-sign method is used to estimate the combinatorial background.
A first order polynomial function is then used to describe the  residual background and a Gaussian function for the signal peak after subtraction of the combinatorial background, as shown in Fig.~\ref{sig_08}. \\

\section{Results}
\begin{figure}[!ht]
\begin{center}
\begin{overpic}[scale=0.45]{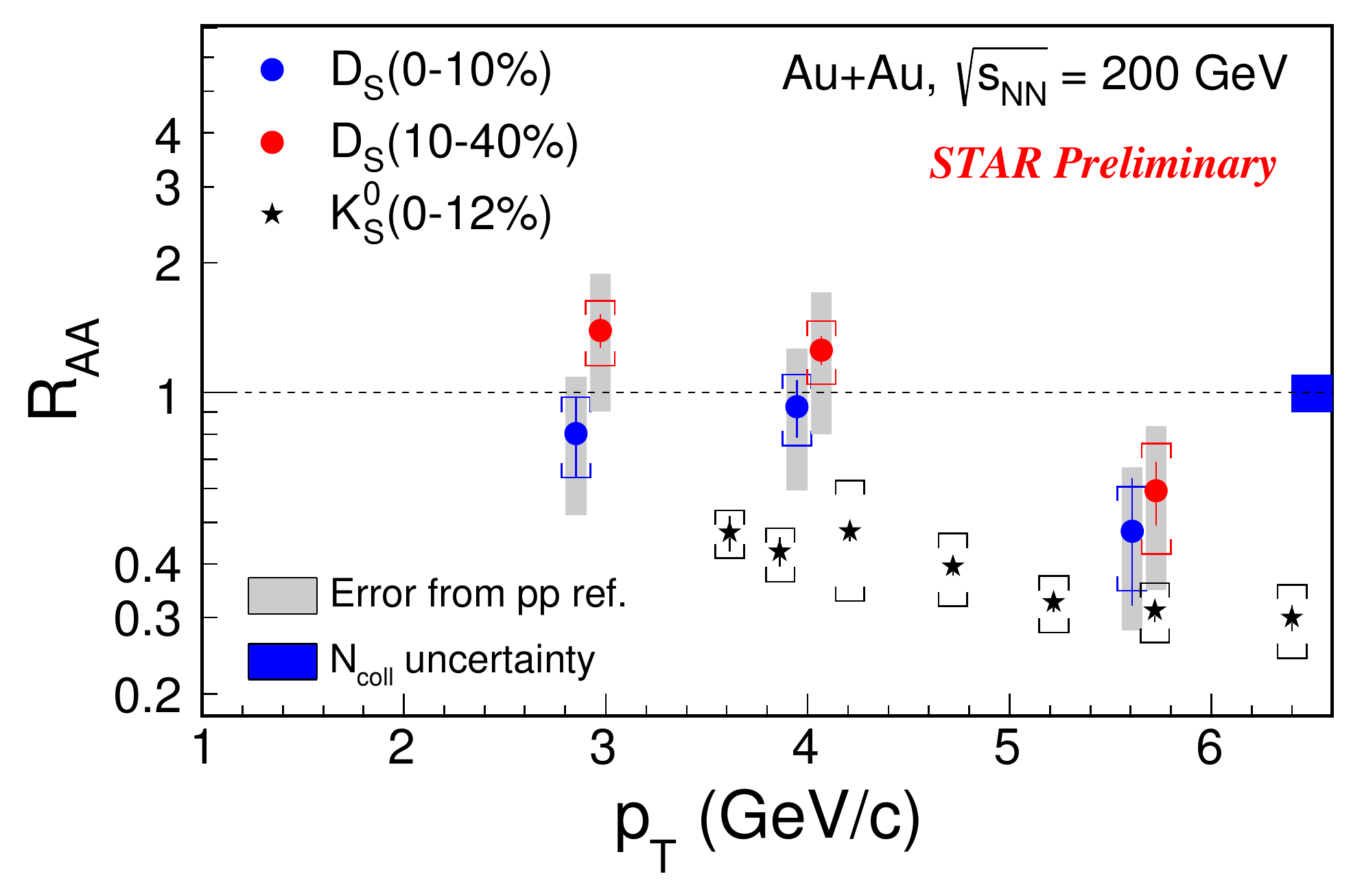}
\end{overpic}
\caption{(Color online)  The  R$_{\rm{AA}}$ of D$_{\rm_{S}}$
   for 0-10$\%$ and  10-40$\%$ central Au+Au collisions at $\sqrt{s_{\rm{NN}}}$  = 200 GeV. Cap symbols are systematic uncertainties  and statistical uncertainties are shown by vertical lines. The R$_{\rm{AA}}$ of K$^{0}_{\rm_{S}}$ are shown as black stars for 0-12$\%$ Au+Au collisions.}
\label{raa}
\end{center}
\end{figure}

The nuclear modification factors (R$_{\rm{AA}}$) of D$_{\rm_{S}}$, as a function of
transverse momentum ($p_{ T}$), for the 0-10$\%$  and 10-40$\%$ most 
central Au+Au collisions at $\sqrt{s_{\rm{NN}}}$ = 200 GeV are shown in Fig.~\ref{raa}. The cap symbols are the systematic uncertainties  and statistical uncertainties are shown by vertical lines. The systematic uncertainties have been
evaluated by using different approaches for background subtraction
and by varying topological cuts. To obtain the R$_{\rm{AA}}$, the charm quark production cross-section measured in p+p collisions by the STAR experiment~\cite{stard0pp} is used together a fragmentation factor f$_{frag}$($c$ $\longrightarrow$ D$_{\rm_{S}}$) = 0.079 $\pm$ 0.004~\cite{ctods}. A Levy fit function is used to obtain the D$_{\rm_{S}}$ yield in p+p collisions  at corresponding measured $p_{T}$ in the Au+Au collisions.  Shaded grey bands represent combined statistical and systematic uncertainties from the p+p reference. 
The R$_{\rm{AA}}$ of K$^{0}_{\rm_{S}}$ are also shown in Fig.~\ref{raa} for 0-12$\%$ Au+Au collisions at $\sqrt{s_{\rm{NN}}}$ = 200 GeV~\cite{k0sraa}. The D$_{\rm_{S}}$ R$_{\rm{AA}}$  is found to be systematically higher than those of  K$^{0}_{S}$. The  R$_{\rm{AA}}$ of D$_{\rm_{S}}$ is consistent with unity at low $p_{T}$ (2.5 <$p_{T}$ < 4 .0 GeV/c) within large uncertainties.  There is an indication of a suppression of high $p_{T}$ (> 5 GeV/c)  D$_{\rm_{S}}$  w.r.t the p+p reference. \\

\begin{figure}[!ht]
\begin{center}
\begin{overpic}[scale=0.36]{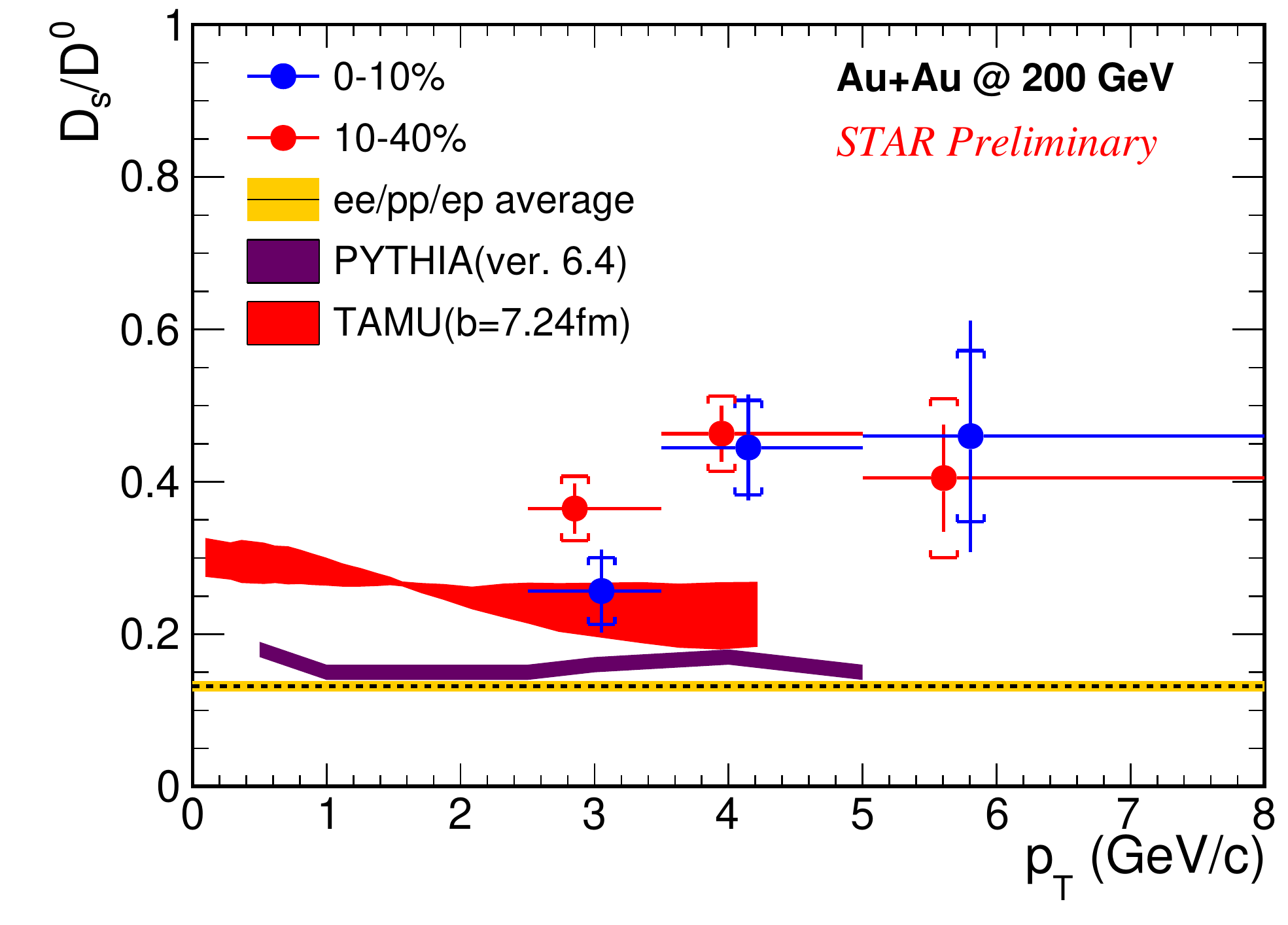}
\end{overpic}
\begin{overpic}[scale=0.39]{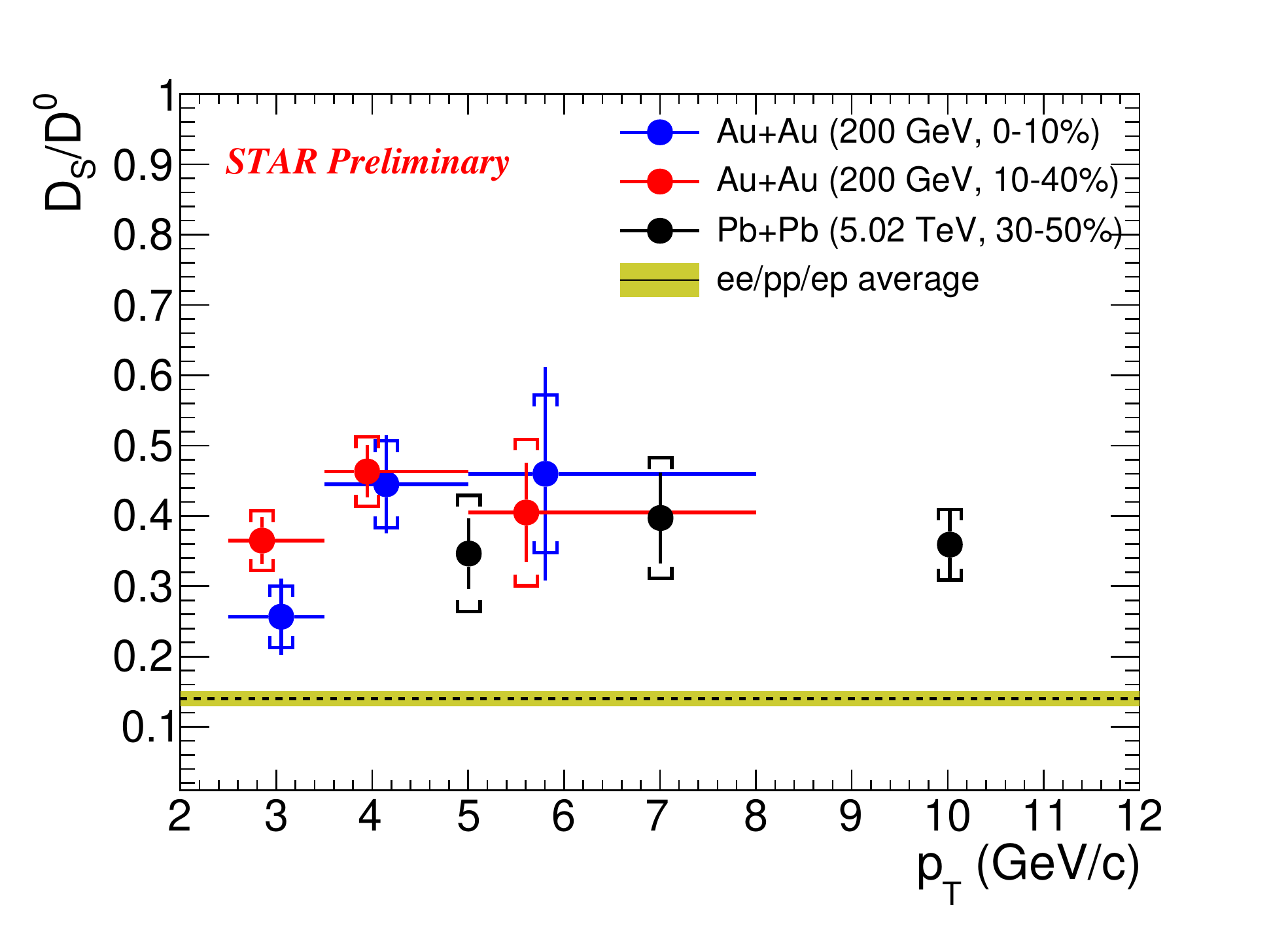}
\end{overpic}
\caption{(Color online)  The ratio of D$_{\rm_{S}}$/D$^{0}$ yields as a function of $p_{T}$  in Au+Au collisions at $\sqrt{s_{\rm{NN}}}$  =200 GeV. Yellow band corresponds to the fragmentation baseline. The magenta and red bands (left panel) represent the PYTHIA  prediction and TAMU model calculation~\cite{rapp}, respectively.   Results from the ALICE collaboration for Pb+Pb collisions at $\sqrt{s_{\rm{NN}}}$  = 5.02 TeV are also shown (right panel). Cap symbols are systematic uncertainties and statistical uncertainties are shown by vertical lines. }
\label{ds_d0_ratio}
\end{center}
\end{figure}

The ratio of D$_{\rm_{S}}$/D$^{0}$ yields, as a function of $p_{T}$, is shown in  Fig.~\ref{ds_d0_ratio}. 
To compare with the D$_{\rm_{S}}$/D$^{0}$ ratio in p+p collisions, we have used the result from the PYTHIA 6.4 Monte Carlo generator~\cite{pythia},
which is shown as a magenta band. The model prediction for Au+Au collisions by the TAMU group is also shown by the red band~\cite{rapp}.
A substantial enhancement in the D$_{\rm_{S}}$/D$^{0}$ ratio in Au+Au collisions w.r.t. the fragmentation baseline (as well as PYTHIA model) is observed.
The TAMU model calculation~\cite{rapp} based on charm quark recombination with enhanced strange quarks also under-predicts data.
We have also compared our results with the ALICE measurement in Pb+Pb collisions at 5.02 TeV~\cite{alice_ds},  as shown in the right panel of Fig.~\ref{ds_d0_ratio}.
The STAR and ALICE results are comparable in the overlapping $p_{T}$ range. \\

\begin{figure}[!ht]
\begin{center}
\begin{overpic}[scale=0.37]{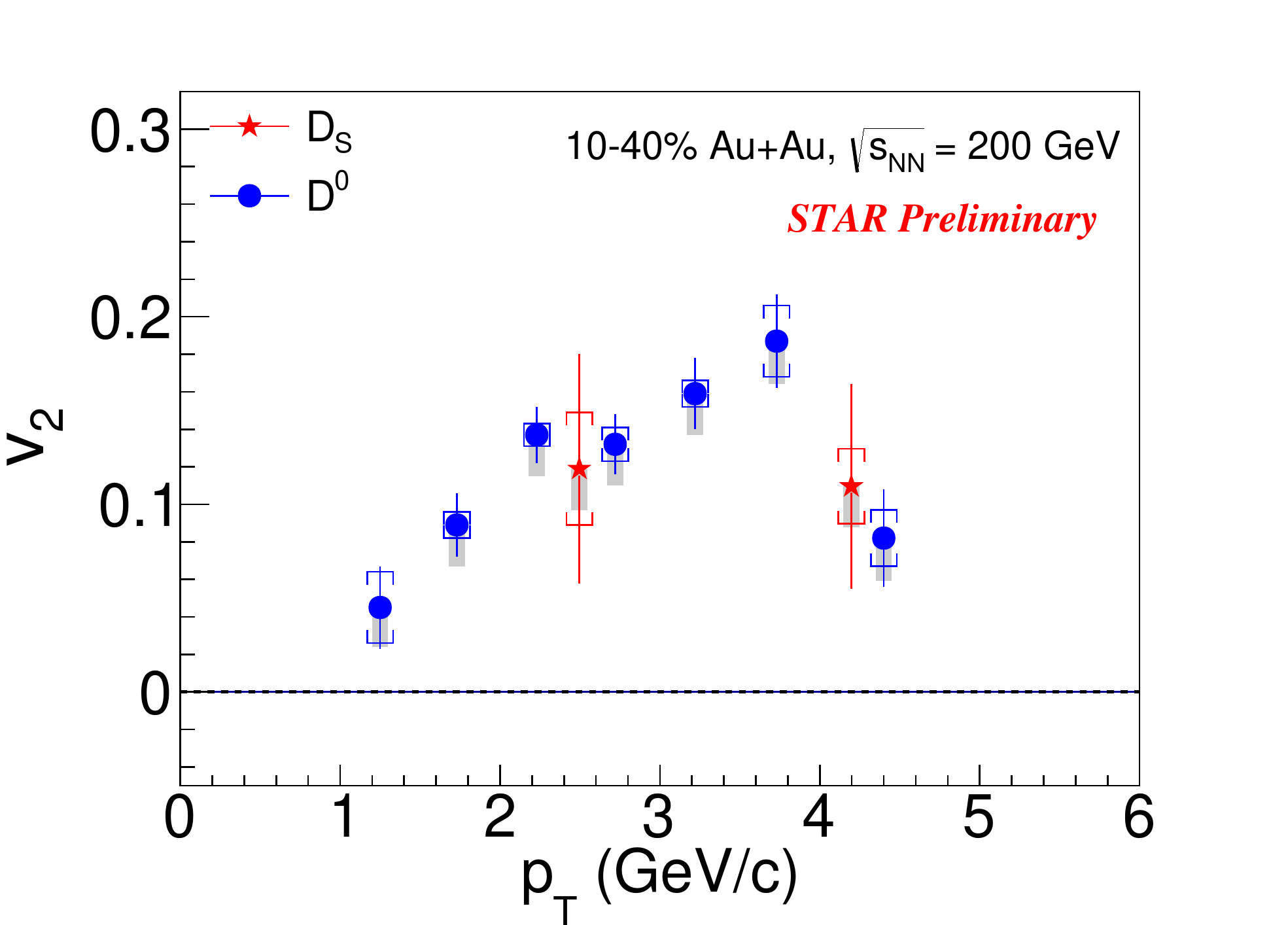}
\end{overpic}
\begin{overpic}[scale=0.37]{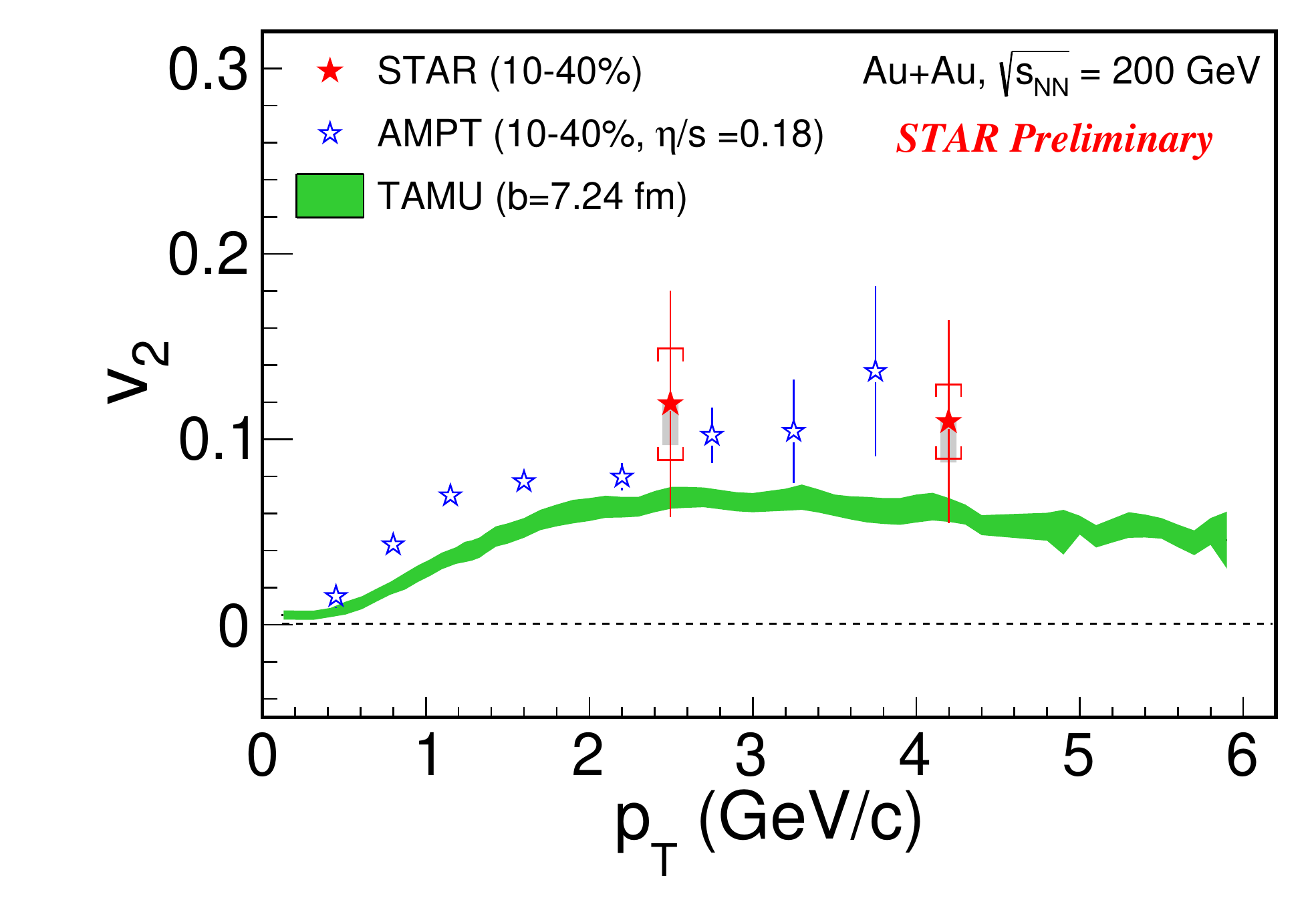}
\end{overpic}
\caption{(Color online)  D$_{\rm_{S}}$ $v_{2}$ as a function of $p_{T}$ in 10-40$\%$ Au+Au collisions at $\sqrt{s_{\rm{NN}}}$  = 200 GeV.  The cap symbols are  systematic uncertainties and statistical uncertainties are shown by vertical lines. Grey bands represent non-flow contribution. }
\label{ds_d0_v2}
\end{center}
\end{figure}

The elliptic flow ($v_{2}$), a measure of the anisotropy in the momentum space, can be used to probe the dynamics of early stages of heavy-ion collisions~\cite{v2intro}. 
The measured D$_{\rm_{S}}$ $v_{2}$, as a function of $p_{T}$, in Au+Au collisions at $\sqrt{s_{\rm{NN}}}$ = 200 GeV is shown in Fig.~\ref{ds_d0_v2}.
These results are obtained by using the event plane method~\cite{v2method,D0v2}. The left panel of Fig.~\ref{ds_d0_v2} shows that the D$_{\rm_{S}}$ $v_{2}$  is non-zero and its magnitude is comparable to the D$^{0}$ $v_{2}$,~\cite{D0v2}. However, the statistical uncertainties are still large. The right panel of Fig.~\ref{ds_d0_v2}  shows a comparison of the D$_{\rm_{S}}$ $v_{2}$ with available model predictions~\cite{rapp,AMPTDsv2}. Both the AMPT  and   TAMU models are in agreement with the data within 1$\sigma$ confidence intervals. In the AMPT model calculation, partonic interactions generate $v_{2}$ and hadronization is done via Dynamic Coalescence Model~\cite{AMPTDsv2}, whereas coupling of the charm quarks to the QGP medium and their subsequent recombination  with equilibrated strange quarks give rise to the $v_{2}$ of D$_{\rm_{S}}$ in the TAMU model~\cite{rapp}.

\section{Summary}
In summary, we present  the nuclear modification factors of D$_{\rm_{S}}$, D$_{\rm_{S}}$/D$^{0}$ ratio and elliptic flow of D$_{\rm_{S}}$ as a function of
transverse momentum in Au+Au collisions at $\sqrt{s_{\rm{NN}}}$ = 200 GeV using the data collected by the STAR experiment in year 2014.  The  R$_{\rm{AA}}$ of D$_{\rm_{S}}$ in 0-10$\%$ and 10-40$\%$ Au+Au collisions are consistent with unity at low $p_{T}$ (2.5 <$p_{T}$ < 4 .0 GeV/c) with large uncertainties and there is an indication of a suppression of high $p_{T}$ (> 5 GeV/c)  D$_{\rm_{S}}$. Production of the light strange K$^{0}_{\rm_{S}}$ meson  is found to be more suppressed compared to that of the heavy D$_{\rm_{S}}$ meson.
We have observed a strong enhancement in the D$_{\rm_{S}}$/D$^{0}$ ratio in Au+Au collisions with respect to that in p+p baseline.  This may indicate that coalescence plays an important role for charm quark hadronization in the QGP.  The elliptic flow of the D$_{\rm_{S}}$  meson is measured in Au+Au collisions (10-40$\%$) at $\sqrt{s_{\rm{NN}}}$ = 200 GeV and  is found to be comparable to the D$^{0}$ $v_{2}$ within  large uncertainties.


\begin{thebibliography}{99}
\bibitem{rapp} M. He, R.J. Fries and R. Rapp,  Phys. Rev. Lett. 110, 112301 (2013).

\bibitem{stard0pp} L. Adamczyk et al. (STAR Collaboration), Phys. Rev. D 86, 72013 (2012).

\bibitem{ctods} M. Lisovyi, et. al., EPJ C 76, 397 (2016).

\bibitem{k0sraa} G. Agakishiev et al. (STAR Collaboration), Phys. Rev. Lett. 108, 72302 (2012).

\bibitem{pythia} T. Sjostrand, S. Mrenna and P. Skands, JHEP 0605:026 (2006).

\bibitem{alice_ds} A. Barbano (for the ALICE Collaboration), Nucl. Phys. A 967, 612 (2017).

\bibitem{v2intro} P. F. Kolb et al., Nucl. Phys. A 715, 653c (2003).
\bibitem{v2method} A. M. Poskanzer and S. A. Voloshin, Phys. Rev. C 58, 1671 (1998).

\bibitem{D0v2} L. Adamczyk et al. (STAR Collaboration), Phys. Rev. Lett. 118, 212301 (2017).
\bibitem{AMPTDsv2} R. Esha et al., JPG 44, 045107 (2017).

\end{thebibliography}
\end{document}